\documentstyle[12pt]{article}

\begin{document}

\begin{flushright}
IMSc/2012/06/14 
\end{flushright} 

\vspace{2mm}

\vspace{2ex}

\begin{center}
{\large \bf General Static (Intersecting) Brane Solutions } \\

\vspace{8ex}

{\large  S. Kalyana Rama}

\vspace{3ex}

Institute of Mathematical Sciences, C. I. T. Campus, 

Tharamani, CHENNAI 600 113, India. 

\vspace{1ex}

email: krama@imsc.res.in \\ 

\end{center}

\vspace{6ex}

\centerline{ABSTRACT}
\begin{quote} 

We present one extra parameter worth of new static brane and
intersecting brane solutions in string/M theory. These general
solutions may be obtained by solving the equations of motion
directly. But, in this letter, we first obtain the general
eleven dimensional vacuum solutions, and then generate the brane
solutions by U duality operations -- namely boosting,
dimensional reduction and uplifting, and S and T dualties. Such
general solutions will be needed in the study of higher
dimensional brane stars and their collapse.

\end{quote}

\vspace{2ex}











\newpage



String/M theory lives in ten/eleven dimensional spacetime and,
upon compactification, is expected to describe our observed
universe. Indeed, among other things, it has given a microscopic
description of extremal and near extremal black holes: Starting
with an effective action, the charged black holes are obtained
from the equations of motion \cite{hs} -- \cite{rt97}. Their
microscopic details in the extremal and near extremal limit are
then well described in terms intersecting brane configurations
in string/M theory and the low energy excitations living on
them. Far-from-extremal cases are also expected to be described
by similar configurations \cite{hm}.

More ambitiously, one may attempt within string/M theory to
describe stable stars, their collapse when sufficiently massive,
and the final collapsed objects -- black holes or otherwise.
Given the success of black hole descriptions in terms of branes,
one may naturally start by taking similar brane systems to be
the constituents of the stars. Then, using appropriate equations
of state and appropriate form of the line element describing
such constituent brane systems, one may develop Oppenheimar --
Volkoff type solutions for the static stars and Oppenheimer --
Snyder type solutions for the collapsing ones. Such studies may
provide insights into cosmic censorship conjecture. They may
also be relevant to Mathur's fuzz ball proposal \cite{fuzz},
according to which there is no horizon and the region within
${\cal O}(1)$ times the Schwarzschild radius is modified and is
filled with string/M theoretic fuzz.

A crucial step in such studies is the matching of the interior
solutions to the exterior ones. A pre requisite for this step
is, clearly, the knowledge of the general exterior solutions.
In four dimensional spherically symetric case, Birkhoff's
theorem guarantees that Schwarzschild and Reissner -- Nordstrom
solutions are the general exterior solutions. To our knowledge,
no similar theorem is known in higher dimensional case with
compact directions. Hence, in the attempt to study string/M
theoretic stars, one first needs to know the general exterior
solutions.

In the ten/eleven dimensional static case, besides the standard
intersecting brane solutions \cite{hs} -- \cite{rt97}, a class
of Kasner type multiparameter solutions are already known
\cite{zz, k05}; but no other class of static solutions. It turns
out that the relevant equations of motion have not been solved
most generally and, thereby, one parameter worth of solutions
are missed. Invariably in all these works, some ansatz is
invoked directly or indirectly while solving, thus limiting the
generality of the solutions.

Invoking no such limiting ansatz, we have obtained the general
static solutions. The existing solutions then follow as
particular cases. In this letter, we present these general
solutions for branes and intersecting branes.

The relevant string/M theory effective action may be written in
the standard notation as
\begin{equation}\label{s0} 
S = \frac{1}{2 \kappa^2} \; \int d^D x \; \sqrt{- g} \; 
\left( {\cal R} - \frac{1}{2} \; (\partial \phi)^2 \;
+ \; \sum_{I = 1}^{\cal N} \frac {e^{\lambda_{p_I} \phi} \; 
(d A_{p_I + 1})^2} {2 (p_I + 2)!} \; \right) \; \; .
\end{equation} 
For string theory branes, $D = 10 \;$, $\phi$ is the dilaton
field, and $\lambda_p = - 1, \; + 1, \;$ or $\frac{3 - p} {2}$
for fundamental strings, NS5 branes, or $D p$ branes
respectively. For M theory branes, $D = 11 \;$ and $\phi$ is
absent. We consider the static case. The appropriate line
element $d s$ for describing the branes and intersecting branes
is then given by
\begin{equation}\label{ds}
d s^2 = - e^{2 \psi} d t^2 
+ \sum_{i = 1}^n e^{2 \lambda^i} (d x^i)^2
+ e^{2 \lambda} d r^2 + e^{2 \sigma} d \Omega_m^2
\end{equation}
where $x^i \;$ denote the toroidal brane directions and
$\Omega_m$ denotes an $m$ dimensional unit sphere. All the
fields are functions of $r \;$ only.

The equations of motion for the fields are derived from the
action $S \;$. Their solutions may be obtained {\em e.g.} by
solving them directly \cite{hs, hm, zz}; or, from extremal
solutions using an algorithm \cite{t96, alg}; or, by performing
U duality operations -- namely boosting, dimensional reduction
and uplifting, and S and T dualties \cite{bho} -- on eleven
dimensional solutions \cite{rt97, k05}. In this letter, we
follow the method of U duality operations \cite{else} : We first
obtain the general solutions for the eleven dimensional vacuum
equations of motion, and then generate the general brane
solutions by performing necessary U duality operations.

The vacuum equations of motion, ${\cal R}_{M N} = 0 \;$, and
equation (\ref{ds}) give
\begin{eqnarray}
\Lambda_r^2 - \psi_r^2 - m \; \sigma_r^2 
- \sum_{i = 1}^n (\lambda_r^i)^2  & = & 
m \; (m - 1) \; e^{2 \lambda - 2 \sigma} \label{rr} \\
\psi_{r r} + (\Lambda_r - \lambda_r) \; \psi_r & = & 0
\label{00} \\
\lambda_{r r}^i + (\Lambda_r - \lambda_r) \; \lambda_r^i & = & 0
\label{ii} \\
\sigma_{r r} + (\Lambda_r - \lambda_r) \; \sigma_r & = & 
(m - 1) \; e^{2 \lambda - 2 \sigma} \label{aa} 
\end{eqnarray}
where $*_r = \frac{d}{d r} * \;$, $\; \Lambda = \psi + \sum_{i =
1}^n \lambda^i + m \sigma$, and we assume that $m \ge 2
\;$. Equations (\ref{00}) and (\ref{ii}) give $\psi = a_0 \; F$,
$\; \lambda^i = a_i \; F$, and
\begin{equation}\label{Fr}
e^{\Lambda - \lambda} \; F_r = (m - 1) \; r_0^{m - 1} 
\end{equation}
where $a_0$, $a_i$, and $r_0$ are integration constants. Using
the freedom in the definition of $r \;$, and trading
$\lambda$ for another function $f \;$, we define
\[
e^{2 \sigma} = r^2 \; e^{2 a_\Omega F} \; \; , \; \; \; 
e^{2 \lambda} = \frac{e^{2 a_\Omega F}}{f} \; \; . 
\]
Thus, the eleven dimensional general vacuum solution is given by
\begin{equation}\label{dsMvac}
d s^2 = - e^{2 a_0 F} d t^2 
+ \sum_{i = 1}^n e^{2 a_i F} (d x^i)^2
+ e^{2 a_\Omega F} \; (d s^\perp_{m + 1})^2 
\end{equation}
where 
\begin{equation}\label{dsm+1}
(d s^\perp_{m + 1})^2 = \frac{d r^2}{f} + r^2 d \Omega_m^2 
\end{equation}
describes the $(m + 1)$ dimensional transverse space. The
constant $a_\Omega$ can be set to zero with no loss of
generality, but its presence will elucidate certain U duality
structure. Let $R = r^{m - 1} \;$, $\; R_0 = r_0^{m - 1} \;$,
\begin{eqnarray}
A = a_0 + \sum_{i = 1}^n a_i + m a_\Omega & , \; \; \; & 
K = a_0^2 + \sum_{i = 1}^n a_i^2 + m a_\Omega^2 - A^2 
\nonumber \\ 
\alpha = \frac{m}{m + 4 (m - 1) K} & , \; \; \; & 
f_0 = 1 - \alpha  \label{alpha} 
\end{eqnarray}
and assume that the $a$'s obey the constraints 
\begin{equation}\label{AKge0}
A - a_\Omega = \frac{1}{2} \; \; , \; \; \; \; 
K \ge 0 \; \; \; \longleftrightarrow \; \; \; 
\alpha \le 1 \; \; . 
\end{equation}
Note that $(a_0, a_i, a_\Omega, r_0)$ and $F(r)$ are defined
only upto a constant multiplicative factor. This factor is fixed
by the first constraint above. The second constraint is imposed
here because we are unable to obtain the complete solution when
$K$ is negative. 

Equations (\ref{rr}), (\ref{aa}), and
(\ref{Fr}) now give
\begin{eqnarray}
f \; (R F_R) & = & 1 - f + \frac{(m - 1) K}{m} \; f \; (R F_R)^2
\label{Reom1} \\
f \; (R F_R) & = & 2 (1 - f) + R f_R \label{Reom2} \\
e^{\frac{F}{2}} \; f^{\frac{1}{2}} \; (R F_R) & = &
\frac{R_0}{R}   \label{Reom3} 
\end{eqnarray}
where $*_R = \frac{d}{d R} * \;$. It follows from these
equations that $f = e^F = 1 - \frac {R_0} {R} \;$ for $K = 0
\;$, and that $f \sim e^F \to 1 - \frac {R_0} {R} \;$ in the
limit $R \to \infty \;$ for $K > 0 \;$. After some algebra, we
get
\begin{equation}\label{RfR}
R f_R = \frac{2 \; (1 - f) \; (f - f_0)}
{f - f_0 + \epsilon \; \sqrt{\alpha \; f \; (f - f_0)}}
\end{equation}
\begin{equation}\label{eF}
e^F = \frac{R_0^2 \; \left( \sqrt{f - f_0} + \epsilon \;
\sqrt{\alpha \; f} \right)^2} {4 \; \alpha \; R^2 \; (1 - f)^2}
\end{equation}
where $\epsilon = \pm 1$ and the square roots are always taken
with postive signs.

As described in \cite{k11}, the qualitative behaviour of $f$ and
$e^F$ for $K > 0 \;$ follows from the above equations :
$\epsilon = + 1$ for $R > R_{min} \;$ and $ = - 1$ for $R <
R_{min} \;$ where $f(R_{min}) = f_0 > 0 \;$; $\; f$ decreases
from $1$ to $f_0$ and then increases to $\infty$ as $R$
decreases from $\infty$ to $R_{min}$ to $0 \;$; and, $e^F$
decreases monotonically from $1$ to $0$, remaining $< f \;$, as
$R$ decreases from $\infty$ to $0 \;$. Note that there is no
horizon.

Let $h = \epsilon \; \sqrt{ \frac{f - f_0}{\alpha}} \;$. Hence,
$h(R_{min}) = 0 \;$ and $h$ decreases monotonically from $1$ to
$- \infty$ as $R$ decreases from $\infty$ to $0 \;$. In terms of
$h$, we get
\begin{equation}\label{feF}
f = 1 - \alpha + \alpha h^2 
\; \; \; , \; \; \; \; 
e^F = \frac{R_0^2}{4 \alpha^2 R^2} \; 
\left( \frac{h + \sqrt{f}} {1 - h^2} \right)^2 
\end{equation}
\begin{equation}\label{RhR}
\frac{d R}{R} = \frac{h + \sqrt{f}} {1 - h^2} \; d h 
\end{equation}
and $- \infty \le h \le 1 \;$. Integrating equation (\ref{RhR})
now gives \cite{Rhcomment}
\begin{equation}\label{r(h)}
\frac{R_{min}}{R} = 
\frac{\sqrt{1 - \alpha} \; \; (1 - h) \; (1 + \sqrt{f})} 
{1 - \alpha + \alpha h + \sqrt{f}} \; \; 
\left( \frac{\sqrt{f} + h \sqrt{\alpha}}{\sqrt{1 - \alpha}}
\right)^{\sqrt{\alpha}} \; \; . 
\end{equation}

We now have the general solutions for $(f, e^F, R)$ as functions
of $h \;$, which depend on two parameters $R_0$ and $K
\;$. Also, since $f \to 1 - \frac {R_0} {R} \;$ in the limit $R
\to \infty \;$, it follows that $h \to 1 - \frac {R_0} {2 \alpha
R} \;$ and, from equation (\ref{r(h)}), that
\[ 
R_{min} = \frac{c(\alpha)}{\alpha} \; R_0 
\; \; \; , \; \; \; \;
c(\alpha) = \frac{1}{2} \; 
\left( 1 + \sqrt{\alpha} \right)^{\frac{1 + \sqrt{\alpha}}{2}}
\; \left( 1 - \sqrt{\alpha} \right)^{\frac{1 - \sqrt{\alpha}}
{2}} \; \; .
\]
Further, one may recover the $K = 0$ solution $f = e^F = 1 -
\frac {R_0} {R} \;$ by formally substituting $\alpha = 1$ in
equations (\ref{feF}) and (\ref{r(h)}).

The general string/M theory brane solutions may now be obtained
by boosting the general vacuum solution and performing U duality
operations. Under a boost {\em e.g.} along $x^n$ direction with
a boost parameter $\beta$, we have
\[
- \; e^{2 a_0 F} d t^2 + e^{2 a_n F} (d x^n)^2
\; \; \longrightarrow \; \; 
- \; \frac{e^{2 a_0 F}}{H} \; d t^2 
+ e^{2 a_n F} \; H \; (d x^n + {\cal A} \; d t)^2
\]
in equation (\ref{dsMvac}) where 
\begin{equation}\label{HA}
H = {\cal C}^2 - e^{2 w F} \; {\cal S}^2 
\; \; , \; \; \; \; 
{\cal A} = \frac{{\cal S} {\cal C}}{H} \; (1 - e^{2 w F}) 
\; \; , 
\end{equation}
$w = a_0 - a_n$, $ \;{\cal S} = Sinh \; \beta$, and ${\cal C} =
Cosh \; \beta \;$. Dimensional reduction along $x^n$ direction
then gives $D0$ branes smeared uniformly along $x^i$ directions,
$i = 1, \cdots, n_s \; (\; = n - 1) \;$. This solution, in
string frame, is given by
\begin{eqnarray}
d s_S^2 & = & - \; H^{- \frac{1}{2}} \; e^{2 b_0 F} d t^2
+ H^{\frac{1}{2}} \; \left( \sum_{i = 1}^{n_s} 
e^{2 b_i F} (d x^i)^2 + e^{2 b_\Omega F} \; d s_{m + 1}^2
\right) \nonumber \\
e^{\phi} & = & H^{\frac{3}{4}} \; e^{b_\phi F} 
\; \; , \; \; \;
A_0 \propto {\cal A} 
\; \; , \; \; \;
w = b_0 - b_\phi \label{D0} 
\end{eqnarray}
where $b_\phi = \frac{3 \; a_n }{2}$ and $b_* = a_* + \frac
{a_n} {2}$ for $* = 0, i, \Omega \;$. The $b$'s must satisfy the
constraints given in equation (\ref{AKge0}) where, by
substituting $a_n = \frac{2 \; b_\phi} {3}$ and $a_* = b_* -
\frac {b_\phi} {3}$ and defining $A_S = b_0 + \sum_{i = 1}^{n_s}
b_i + m b_\Omega \;$, we have
\begin{eqnarray}
A - a_\Omega & = &  A_S - 2 b_\phi - b_\Omega
\label{Ab} \\
K & = & b_0^2 + \sum_{i = 1}^{n_s} b_i^2 + m b_\Omega^2 
- (A_S - 2 b_\phi)^2 \; \; . \label{Kb}
\end{eqnarray}

The expressions for $(A - a_\Omega)$ and $K$, written in terms
of $b$'s, remain invariant under $T$ and $S$ dualities; and,
written in terms of $a$'s, remain invariant under further
dimensional uplifting \cite{check}. Hence, in the following, we
will not distinguish between the original and transformed $a$'s
and $b$'s and, further, assume that all the $a$'s and $b$'s
satisfy the constraints in equation (\ref{AKge0}).

We now proceed to obtain general brane solutions. T dualising
$D0$ branes along $(x^1, \cdots, x^p)$ directions gives $Dp$
branes. S dualising $D1$ and $D5$ branes gives fundamental
strings and $NS5$ branes. Lifting $D2$ and $D4$ branes to eleven
dimensions gives $M2$ and $M5$ branes. Performing these
operations using the formulas given in \cite{bho}, it follows
that the general string theory $p$ brane solutions may all be
written, in string frame, as
\begin{eqnarray}
d s_S^2 & = & H^{A_\perp - 1} \; \left( - e^{2 b_0 F} d t^2
+ \sum_{i = 1}^{p} e^{2 b_i F} (d x^i)^2 \right) 
+ H^{A_\perp} \; e^{2 b_\Omega F} \; (d s^\perp_{9 - p})^2
\nonumber \\
e^{\phi} & = & H^{\frac{\lambda_p}{2}} \; e^{b_\phi F}
\; \; , \; \; \; 
w = b_0 + \sum_{i = 1}^p b_i - 2 A_\perp b_\phi \label{Dp}
\end{eqnarray}
where $\lambda_p = - 1, \; + 1, \; \frac{3 - p} {2}$ and
$A_\perp = 0, \; 1, \; \frac{1}{2} \;$ for fundamental strings,
NS5 branes, and $D p$ branes respectively. The general M theory
$p$ brane solutions may be written as
\begin{eqnarray}
d s^2 & = & H^{\frac{p - 8}{9}} \; \left( - e^{2 a_0 F} d t^2 
+ \sum_{i = 1}^{p} e^{2 a_i F} (d x^i)^2 \right) 
+ H^{\frac{p + 1}{9}} \; e^{2 a_\Omega F} \; 
(d s^\perp_{10 - p})^2 \nonumber \\
w & = &  a_0 + \sum_{i = 1}^p a_i 
\; \; . \label{Mp}
\end{eqnarray}

In these solutions, the non zero $(p +1)-$form gauge field
component $A_{0 1 \cdots p} \propto {\cal A} \;$; the $p$ brane
charge $\propto w R_0 {\cal S} {\cal C} \;$; and, the functions
$H$ and ${\cal A}$ are given by equations (\ref{HA}), with $w$
given as above. Note that the behaviour of $H$ follows easily
from that of $e^F \;$: For example, for $w > 0$, the function
$H$ increases monotonically from $1$ to ${\cal C}^2$ as $R$
decreases from $\infty$ to $R_s \;$ where $R_s = R_0$ if $K = 0$
and $R_s = 0$ if $K > 0 \;$. Note that there is no horizon in
the $K > 0$ case.

Note also that one may formally take Maldacena's decoupling
limit. Then $H$ becomes $(1 - e^{2 w F}) \; {\cal S}^2 \;$.
Since $e^F \to 1 - \frac {R_0} {R} \;$ as $R \to \infty \;$, it
follows that the resulting spacetimes for $D3$, $M2$, and $M5$
branes are asymptotically $AdS_{p + 2} \times S^q$ where $(p, q)
= (3, 5), (2, 7)$, and $(5, 4)$ respectively. But, now, the
interiors of these spacetimes are generically modified. For
example, for $K = 0$ in the $D3$ brane case but with $b_0 = b_i$
and $\; b_\Omega = \frac {b_\phi} {4}$, it can be shown in the
limit $R \to R_0$ that the modified interior spacetime behaves
the same way as that found in \cite{99}.

The general ${\cal N}$ intersecting brane solutions may be
generated by performing ${\cal N}$ number of boosts and other U
duality operations appropriately. For example, the general
solution for three stacks of intersecting M2 branes is given by
\begin{eqnarray}
d s^2 & = & - \; 
\left( H_2 H_{2'} H_{2''} \right)^{- \frac{2}{3}}
\; e^{2 a_0 F} \; d t^2 
+ \left( \frac {H_{2'} H_{2''}} {H_2^2} \right)^{\frac{1}{3}} \;
\sum_{i = 1}^2 e^{2 a_i F} \; (d x^i)^2 \nonumber \\
& + & \left( \frac {H_{2} H_{2''}} {H_{2'}^2}
\right)^{\frac{1}{3}} \; \sum_{i = 3}^4 e^{2 a_i F} \; (d x^i)^2
+ \left( \frac {H_{2} H_{2'}} {H_{2''}^2} \right)^{\frac{1}{3}}
\; \sum_{i = 5}^6 e^{2 a_i F} \; (d x^i)^2 \nonumber \\
& & \nonumber \\
& + & \left( H_2 H_{2'} H_{2''} \right)^{\frac{1}{3}} \;
e^{2 a_\Omega F} \; (d s^\perp_4)^2 \nonumber \\
w_2 & = & a_0 + \sum_{i = 1}^2 a_i 
\; \; , \; \; \; \; 
w_{2'} = a_0 + \sum_{i = 3}^4 a_i 
\; \; , \; \; \; \; 
w_{2''} = a_0 + \sum_{i = 5}^6 a_i  \label{22'2''} 
\end{eqnarray}
where, for $I = (2, 2', 2'') \;$,  
\begin{equation}\label{HAI}
H_I = {\cal C}^2_I - e^{2 w_I F} \; {\cal S}^2_I 
\; \; , \; \; \; \; 
{\cal A_I} = \frac{{\cal S}_I {\cal C}_I}{H_I} \; 
(1 - e^{2 w_I F}) \; \; , 
\end{equation}
${\cal S}_I = Sinh \; \beta_I$, $\; {\cal C}_I = Cosh \;
\beta_I$, and $\beta_I$ are boost parameters. The non zero
$3-$form gauge field components $\propto {\cal A}_I \;$ and the
brane charges $\propto w_I R_0 {\cal S}_I {\cal C}_I \;$.

It turns out, similarly as in \cite{alg}, that the general brane
and intersecting brane solutions may all be written down
algorithmically as follows. For M theory case: 

\begin{itemize}

\item

Start with the extremal solution \cite{t96}.

\item

In $d s^2$, attach the function $f$ to the $d r^2$ term and the
$e^{2 a_* F}$ factors to the other terms as in the above
solutions. The $a$'s must satisfy the constraints in equation
(\ref{AKge0}). The functions $f$ and $e^F$ are given by
equations (\ref{feF}) where $h(r)$ is implicitly given by
equation (\ref{r(h)}).

\item

Replace the harmonic and the gauge field functions of the
extremal solution by the functions $(H_I, {\cal A}_I)$ given in
equation (\ref{HAI}). The $\beta$'s will be determined by the
brane charges. The $w$'s for each stack of branes $= \sum_{wld
\; vol} a_* \;$, {\em i.e.}  the sum of $a$'s on its
worldvolume.

\end{itemize}

\noindent
For string theory case: In the above steps, $a$'s are replaced
by $b$'s; an $e^{b_\phi F}$ factor is attached for dilaton; and
the expression for $w$ has an additional term $= - 2 A_\perp
b_\phi \;$.

The existing brane and intersecting brane solutions now follow
as particular cases of our general solutions : The standard
solutions \cite{hs} -- \cite{rt97} follow for $K = 0$, $a_0 =
b_0 = \frac {1} {2}$, and $a_\times = b_\times = 0$ if $\times
\ne 0 \;$. Then $w_I = \frac{1}{2} \;$ and
\[
f = e^F = 1 - \frac {r_0^{m - 1}} {r^{m - 1}} 
\; \; , \; \; \; 
H_I = 1 + \frac{r_0^{m - 1} \; {\cal S}^2_I} {r^{m - 1}} 
\; \; . 
\]
The Kasner type solutions \cite{zz, k05} follow for $K = 0 \;$
where, now, the $a$'s and $b$'s satisfy the constraints in
equation (\ref{AKge0}) and are otherwise arbitrary. Then $(f,
e^F)$ are unchanged and $(w_I, H_I)$ depend on the values of
$a$'s and $b$'s.

Also, note that $r = r_0$ is a regular horizon for the standard
solutions. For other solutions, namely the Kasner type ones
where $K = 0$ and the general ones where $K > 0$, the tidal
forces can be shown \cite{zz, k05, k11} to diverge as $r \to
r_s$ where $r_s = r_0$ if $K = 0$ and $r_s = 0$ if $K > 0 \;$.
We take this divergence as implying that there is a naked
curvature singularity at $r_s \;$.

These singular solutions may simply be ignored by invoking
cosmic censorship conjecture and declaring them unphysical. On
the other hand, and more constructively, one may use these
general solutions and study static brane stars and their
collapse. This is likely to provide insights into cosmic
censorship. It is also possible, but perhaps far-fetched within
the present effective action framework, that the singularities
at $r_s$ may be resolved by string/M theoretic effects. Along a
different line, one may assume that the present general
solutions for $m = 2$ describe the exterior of the actual stars
in our universe and, as is being done in several recent works
{\em e.g.} \cite{panda, pankaj}, study their experimental
signatures.





\begin{thebibliography}{999}


\bibitem{hs}
G.~T.~Horowitz and A.~Strominger,
Nucl.\ Phys.\ B {\bf 360}, 197 (1991); 
\\
M.~J.~Duff, R.~R.~Khuri and J.~X.~Lu, \\
Phys.\ Rept.\  {\bf 259}, 213 (1995)
[hep-th/9412184]; 
\\
N.~Ohta,
Phys.\ Lett.\ B {\bf 403}, 218 (1997)
[hep-th/9702164]. 

\bibitem{hm}
G.~T.~Horowitz, J.~M.~Maldacena and A.~Strominger, \\
Phys.\ Lett.\ B {\bf 383}, 151 (1996)
[hep-th/9603109]; 
\\
G.~T.~Horowitz, D.~A.~Lowe and J.~M.~Maldacena, \\
Phys.\ Rev.\ Lett.\  {\bf 77}, 430 (1996)
[hep-th/9603195].

\bibitem{t96}
A.~A.~Tseytlin,
Nucl.\ Phys.\ B {\bf 475}, 149 (1996)
[hep-th/9604035]; 
\\
I.~R.~Klebanov and A.~A.~Tseytlin, \\
Nucl.\ Phys.\ B {\bf 475}, 179 (1996)
[hep-th/9604166];   
\\
J.~P.~Gauntlett, D.~A.~Kastor and J.~H.~Traschen, \\
Nucl.\ Phys.\ B {\bf 478}, 544 (1996)
[hep-th/9604179];   
\\
A.~A.~Tseytlin,  
Nucl.\ Phys.\ B {\bf 487}, 141 (1997)
[hep-th/9609212].   

\bibitem{alg}
M.~Cvetic and A.~A.~Tseytlin, \\
Nucl.\ Phys.\ B {\bf 478}, 181 (1996)
[hep-th/9606033].

\bibitem{rt97}
J.~G.~Russo and A.~A.~Tseytlin, \\
Nucl.\ Phys.\ B {\bf 490}, 121 (1997)
[hep-th/9611047];   
\\
A.~A.~Tseytlin,
Phys.\ Lett.\ B {\bf 395}, 24 (1997)
[hep-th/9611111]; 
\\
N.~Ohta and T.~Shimizu, \\
Int.\ J.\ Mod.\ Phys.\ A {\bf 13}, 1305 (1998)
[hep-th/9701095];   
\\
S.~R.~Das, S.~D.~Mathur, S.~Kalyana Rama and P.~Ramadevi, \\
Nucl.\ Phys.\ B {\bf 527}, 187 (1998)
[hep-th/9711003].  

\bibitem{fuzz}
S.~D.~Mathur,
Fortsch.\ Phys.\  {\bf 53}, 793 (2005)
[hep-th/0502050];   
\\
K.~Skenderis and M.~Taylor, \\
Phys.\ Rept.\  {\bf 467}, 117 (2008)
[arXiv:0804.0552 [hep-th]]; 
\\
B.~D.~Chowdhury and A.~Virmani,
arXiv:1001.1444 [hep-th].

\bibitem{zz}
B.~Zhou, C.~-J.~Zhu,
[hep-th/9905146]; 
\\
P.~Brax, G.~Mandal, Y.~Oz, \\
Phys.\ Rev.\  {\bf D63}, 064008 (2001), 
[hep-th/0005242];
\\
J.~D.~Edelstein, J.~Mas,
JHEP {\bf 06}, 015 (2004), 
[hep-th/0403179];
\\
Y.~-G.~Miao, N.~Ohta, \\
Phys.\ Lett.\  {\bf B594}, 218 (2004), 
[hep-th/0404082]; 
\\
S.~Kobayashi, T.~Asakawa, S.~Matsuura, \\
Mod.\ Phys.\ Lett.\  {\bf A20}, 1119 (2005), 
[hep-th/0409044];
\\
C.~-M.~Chen, D.~V.~Gal'tsov and N.~Ohta, \\
Phys.\ Rev.\ D {\bf 72}, 044029 (2005)  
[hep-th/0506216].  

\bibitem{k05}
S.~Kalyana~Rama,
[hep-th/0503058].

\bibitem{bho}
E.~Bergshoeff, C.~M.~Hull and T.~Ortin, \\ 
Nucl.\ Phys.\ B {\bf 451}, 547 (1995)
[hep-th/9504081]. 

\bibitem{else}
We have also solved the equations of motion directly, but the
details are quite involved. We will present them elsewhere.

\bibitem{k11}
S.~Kalyana~Rama,
arXiv:1111.1897 [hep-th].

\bibitem{Rhcomment}
Expand $\frac{h + \sqrt{f}} {1 - h^2}$ into partial fractions,
each of which integrates to $ln$ or $ArcSinh$ functions.
Repeated use of the formulas $ArcSinh x = ln (x + \sqrt{1 +
x^2})$ and $a \pm \sqrt{b} = \frac {a^2 - b} {a \mp \sqrt{b}}$
then leads to equation (\ref{r(h)}), which indeed satisfies
equation (\ref{RhR}) as may be checked by direct substitution.

\bibitem{check}
This follows because under a T duality {\em e.g.}  along $x^1$
direction, the $b$'s transform as $\left( b_\phi, \; b_1, \;
b_{* \ne 1} \right) \rightarrow \left(b_\phi - b_1, \; - b_1, \;
b_{* \ne 1} \right)$ whereas, under an S duality, they transform
as $( b_\phi, \; b_* ) \rightarrow ( - b_\phi, \; b_* - \frac
{b_\phi} {2}) \;$. Invariance under further dimensional
uplifting follows by reversing the steps which led to equations
(\ref{Ab}) and (\ref{Kb}).

\bibitem{99}
A.~Kehagias and K.~Sfetsos, \\
Phys.\ Lett.\ B {\bf 454}, 270 (1999)
[hep-th/9902125]; 
\\
S.~S.~Gubser,
hep-th/9902155; 
\\
L.~Girardello, M.~Petrini, M.~Porrati and A.~Zaffaroni, \\
JHEP {\bf 05}, 026 (1999)
[hep-th/9903026].

\bibitem{panda}
H.~Liu and J.~M.~Overduin,  \\
Astrophys.\ J.\  {\bf 538}, 386 (2000)
[gr-qc/0003034];
\\
A.~Bhattacharjee, A.~Das, L.~Greenwood and S.~Panda, \\
Int.\ J.\ Mod.\ Phys.\ D {\bf 21}, 1250056 (2012)
arXiv:1112.0887 [hep-th].

\bibitem{pankaj}
K.~S.~Virbhadra and G.~F.~R.~Ellis,
Phys.\ Rev.\ D {\bf 65}, 103004 (2002); 
\\
Z.~Kovacs and T.~Harko, \\
Phys.\ Rev.\ D {\bf 82}, 124047 (2010)
[arXiv:1011.4127 [gr-qc]];
\\
A.~N.~Chowdhury, M.~Patil, D.~Malafarina and P.~S.~Joshi, \\
Phys.\ Rev.\ D {\bf 85}, 104031 (2012)
arXiv:1112.2522 [gr-qc]; 
\\
M.~Patil and P.~S.~Joshi, \\
Phys.\ Rev.\ D {\bf 85}, 104014 (2012)
[arXiv:1112.2525 [gr-qc]].  

\end{thebibliography}
\end{document}